\documentclass[english,aip,apl,reprint]{revtex4-1}
\usepackage{helvet}
\usepackage[T1]{fontenc}
\usepackage[latin9]{inputenc}
\usepackage{amsbsy}
\usepackage{amssymb}
\usepackage{graphicx}
\usepackage{esint}
\usepackage{babel}
\begin{document}

\title{Piezoelectric field, exciton lifetime, and cathodoluminescence intensity
at threading dislocations in GaN\{0001\}}

\author{Vladimir M. Kaganer}

\address{Paul-Drude-Institut für Festkörperelektronik, Hausvogteiplatz 5\textendash 7,
10117 Berlin, Germany}

\author{Karl K. Sabelfeld}

\affiliation{Institute of Computational Mathematics and Mathematical Geophysics,
Russian Academy of Sciences, Lavrentiev Prosp.~6, 630090 Novosibirsk,
Russia}

\author{Oliver Brandt}

\address{Paul-Drude-Institut für Festkörperelektronik, Hausvogteiplatz 5\textendash 7,
10117 Berlin, Germany}

\begin{abstract}
The strain field of a dislocation emerging at a free surface is partially
relaxed to ensure stress free boundary conditions. We show that this
relaxation strain at the outcrop of edge threading dislocations in
GaN\{0001\} gives rise to a piezoelectric volume charge. The electric
field produced by this charge distribution is strong enough to dissociate
free excitons at distances over 100~nm from the dislocation line.
We evaluate the impact of this effect on cathodoluminescence images
of dislocations.
\end{abstract}

\maketitle

Cathodoluminescence (CL) maps of GaN\{0001\} surfaces reveal threading
dislocations as dark spots, directly demonstrating that they act as
centers of nonradiative recombination of excitons.\cite{rosner97,sugahara98,speck99,pauc06}
The nonradiative recombination process is commonly presumed to take
place at the dislocation core. The size of the dark spots, typically
some 100~nm in diameter, is understood to be a result of exciton
diffusion, and intensity profiles across the dislocation have been
used frequently to extract quantitative information on the exciton
diffusion length.\cite{rosner97,sugahara98,speck99,shmidt02,nakaij05,pauc06,yakimov07,ino08,yakimov10,yakimov15,sabelfeld17CL}

Wurtzite GaN is a pyroelectric (and thus piezoelectric) crystal. Uniform
strain produces a piezoelectric polarization, and strain variations
generate a piezoelectric field that may dissociate excitons. In bulk
GaN, neither $a$-type edge nor $c$-type screw dislocations with
the line direction along $\left\langle 0001\right\rangle $ cause
piezoelectric fields,\cite{smirnova74,shi99} in contrast to dislocations
with other line directions. When these dislocations reach a surface,
a polarization charge is created due to the discontinuity of the piezoelectric
polarization as well as of the spontaneous polarization. In a previous
work, this charge was presumed to be screened by free charges in surface
states, and it was concluded that the piezoelectric fields of threading
dislocations have a minimal effect on the electric and optical properties
of GaN\{0001\}.\cite{shi99} 

These considerations were based on the strain field of a straight
dislocation in bulk GaN, which does not vary along the $c$ axis.
However, the strain field of the dislocation changes in the vicinity
of the surface to produce a stress-free boundary.\cite{yoffe61,lothe92}
The resulting three-dimensional electric field at the dislocation
outcrop in GaN\{0001\} was calculated numerically by a finite element
approach\cite{taupin14} as an illustration of the computational method,
but the consequences of this field were not considered.

In the present work, we calculate the distribution of the piezoelectric
field that arises due to elastic strain relaxation at the surface
around the outcrops of threading dislocations in GaN\{0001\}. This
field drastically reduces the exciton lifetime close to the dislocation
outcrop. We calculate the CL image of the dislocation in the absence
of exciton diffusion, with exciton dissociation as the only reason
for a variation of the CL signal near the dislocation. The widths
of the CL intensity profiles thus obtained can approach the widths
of the profiles observed experimentally.

A rigorous calculation of the piezoelectric field requires to take
into account both the direct (polarization caused by strain) and the
converse (stress caused by electric field) piezoelectric effects,
necessitating a selfconsistent solution of the coupled equations of
elasticity and electrostatics.\cite{nye57,landau:electrodyn,auld73,nowacki07,voon11}
The relative contributions of the two effects and the need for the
rigorous solution is examined in the following by order of magnitude
estimates. In the next two paragraphs, the orientational dependencies
and the indices of the respective tensors are omitted and the sign
$\sim$ means ``on the order of''. 

The direct effect results in a polarization $P\sim e\varepsilon$,
where $e$ and $\varepsilon$ are the characteristic magnitudes of
the piezoelectric constants and the strain, respectively. The converse
effect produces a stress $\sigma\sim eE$ with the electric field
$E$ and the same piezoelectric constants $e$ as for the direct effect.
This stress is added to the stress $\sigma\sim C\varepsilon$ caused
by elastic strain (here $C$ is the characteristic magnitude of the
elastic moduli). As a first approximation, the electric field can
be taken to be equal to that induced by the direct effect, $E\sim P/\kappa_{0}\kappa$,
where $\kappa_{0}$ is the vacuum permittivity and $\kappa=9.5$ is
the relative permittivity of GaN. Substituting this field results
in a stress of $\sigma\sim(1+K^{2})C\varepsilon$, where $K^{2}=e^{2}/\kappa_{0}\kappa C$
is the electromechanical coupling constant (in a precise definition,
this constant is expressed through the components of the involved
tensors and depends on directions in the crystal\cite{auld73}). This
dimensionless quantity controls the ratio of the converse to direct
piezoelectric effects. Taking for the estimate $e\sim e_{0}/a^{2}\sim1$~C~m$^{-2}$
with the elementary charge $e_{0}$ and the lattice constant $a$,
and $C=200$~GPa, we arrive at $K^{2}=0.06$. This estimate agrees
with the impact of the converse piezoelectric effect in planar heteroepitaxial
structures.\cite{voon11} 

In addition to the piezoelectric effect, the spatially varying strain
field of a dislocation gives rise to a polarization due to the flexoelectric
effect. This effect takes place in all dielectric materials and induces
a polarization $P\sim f\mathrm{\thinspace grad}\,\varepsilon$ (see
Refs.~\onlinecite{maranganti09,hong13,yudin13} and references therein).
With the characteristic magnitude of the flexoelectric constants $f\sim e_{0}/a$
and the dislocation strain $\varepsilon\sim b/2\pi r_{\perp}$ (where
$b$ is the Burgers vector and $r_{\perp}$ is the distance from the
dislocation line), the ratio of the flexoelectric and piezoelectric
effects is $\sim a/r_{\perp}$. Hence, at distances from the dislocation
large in comparison with the atomic distances, the flexoelectric effect
can be neglected.

Given the fact that the converse piezoelectric and the flexoelectric
effects result in corrections of only a few percent, much less than
the uncertainty introduced by the low accuracy with which the piezoelectric
constants are known, we do not include them in the following calculations.
In this case, the piezoelectric polarization $\boldsymbol{P}$ caused
by strain in a crystal with the wurtzite structure has the components
\begin{eqnarray}
P_{x} & = & 2e_{15}\varepsilon_{xz},\,\,\,\,\,P_{y}=2e_{15}\varepsilon_{yz},\nonumber \\
P_{z} & = & e_{31}(\varepsilon_{xx}+\varepsilon_{yy})+e_{33}\varepsilon_{zz},\label{eq:1}
\end{eqnarray}
where $e_{13}$, $e_{33}$, and $e_{15}$ are the piezoelectric constants
of the wurtzite structure and $\varepsilon_{ij}$ are the components
of the strain tensor. The $z$ axis is directed along the {[}0001{]}
direction. Using analytical expressions for the displacement field
of an edge dislocation normal to the surface\cite{yoffe61,lothe92}
and calculating the corresponding components of the strain tensor,
the components of the polarization vector, and the polarization charge
density $\varrho(\boldsymbol{r})=-\mathrm{div}\,\boldsymbol{P}$,
we obtain
\begin{equation}
\varrho(\boldsymbol{r})=y\left(\frac{f_{1}}{r^{3}}+\frac{f_{2}z^{2}}{r^{5}}\right),\label{eq:2}
\end{equation}
where $r=\left(x^{2}+y^{2}+z^{2}\right)^{1/2}$ and
\begin{eqnarray}
f_{1} & = & \frac{b}{\pi}\frac{\nu}{1-\nu}\left[\nu e_{33}-(1-\nu)e_{31}\right],\label{eq:3}\\
f_{2} & = & \frac{3b}{2\pi}\frac{\nu}{1-\nu}\left[e_{31}+2e_{15}-e_{33}\right],\label{eq:4}
\end{eqnarray}
with the Poisson ratio $\nu$. We restrict our calculations to the
elastically isotropic case. The expressions for the displacement field
of an edge dislocation far from the surface, written for an isotropic
medium, remain valid for an edge dislocation in the elastically anisotropic
hexagonal crystal with the dislocation line along the $c$ axis, if
the Poisson ratio is taken as $\nu=C_{12}/(C_{11}+C_{12})$, where
the $C_{ij}$ are the elastic moduli.\cite{belov92} With the elastic
moduli of GaN,\cite{polian96} we obtain $\nu=0.27$. We use this
value also for the relaxation displacement field. 

For the piezoelectric constants of GaN, we take the values $e_{31}=-0.527$~C~m$^{-2}$
and $e_{33}=0.895$~C~m$^{-2}$.\cite{winkelnkemper06} For $e_{15}$
we assume, as proposed in Ref.~\onlinecite{schulz11}, $e_{15}=e_{31}$.
With the Burgers vector of an \emph{a}-type edge dislocation $b=0.319$~nm,
we obtain the values $f_{1}=0.235\times10^{-10}$~C~m$^{-1}$ and
$f_{2}=-1.39\times10^{-10}$~C~m$^{-1}$.

We have performed the same calculation for the charge density for
a $c$-type screw dislocation. The strain relaxation at the free surface
gives rise to a non-zero piezoelectric polarization $\boldsymbol{P}$,
but the charge density $\varrho(\boldsymbol{r})=-\mathrm{div}\,\boldsymbol{P}$
is equal to zero. Hence, screw threading dislocations do not generate
a piezoelectric field. Since the equations of both elasticity and
electrostatics are linear, mixed ($a+c$-type) dislocations produce
the same piezoelectric field as $a$-type edge dislocations.

To obtain the piezoelectirc field associated with the charge density
(\ref{eq:2}), we consider first the case of undoped GaN, so that
free charges in the bulk are absent and the electric potential $V(\boldsymbol{r})$
is the solution of the Poisson equation
\begin{equation}
\nabla^{2}V=-\frac{\varrho}{\kappa_{0}\kappa}.\label{eq:5}
\end{equation}
We assume that the surface charges are screened by free charges, so
that the boundary condition for Eq.~(\ref{eq:5}) reads $V\left|_{z=0}\right.=0$.

The solution of Eq.~(\ref{eq:5}) for a point charge with the equipotential
boundary condition $V\left|_{z=0}\right.=0$ is the Coulomb potential
$1/4\pi\kappa_{0}\kappa r$ plus that of the image (of the opposite
sign) with respect to the surface $z=0$. Integrating the point charge
solution with the charge distribution (\ref{eq:2}), we arrive at
the potential

\begin{equation}
V(x,y,z)=V_{0}\frac{yz}{r_{\perp}^{2}}\left[\left(1-\frac{z}{r}\right)+\frac{f_{2}}{6f_{1}}\left(1-\frac{z^{3}}{r^{3}}\right)\right],\label{eq:6}
\end{equation}
where $V_{0}=f_{1}/2\kappa_{0}\kappa$ and $r_{\perp}=\left(x^{2}+y^{2}\right)^{1/2}$.
We do not present the details of the derivation since the result can
be directly verified by differentiating expression (\ref{eq:6}) and
checking that Eq.~(\ref{eq:5}) is satisfied with the charge density
(\ref{eq:2}).

\begin{figure}
\includegraphics[width=1\columnwidth]{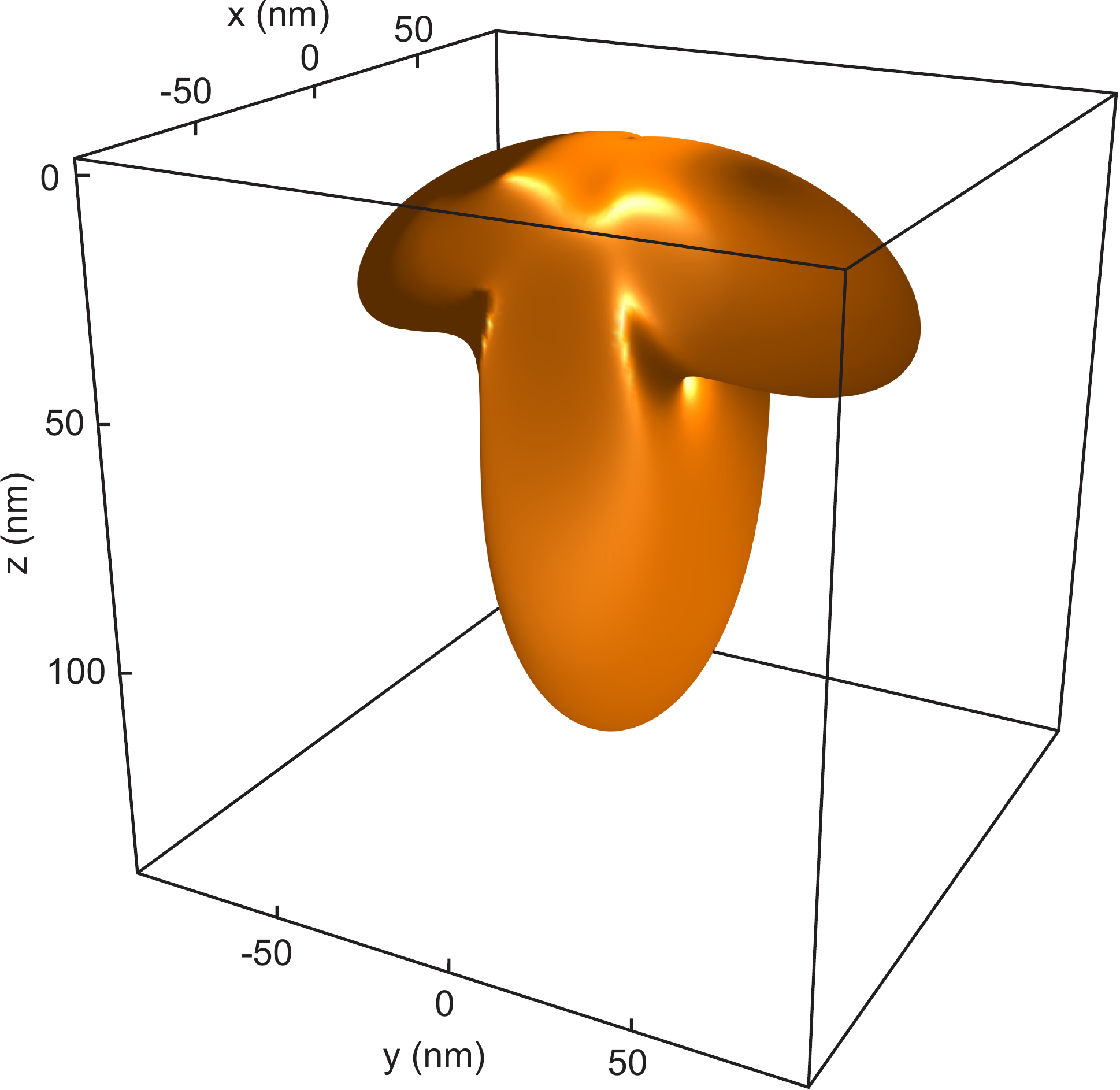}

\caption{Electric field distribution $E=|\boldsymbol{E}|$ around the outcrop
of an edge threading dislocation. The figure shows an iso-field surface
at $E=10$~kV/cm. The $xy$ plane at $z=0$~corresponds to the GaN\{0001\}
surface.}

\label{fig:ElectricField}
\end{figure}

With the numerical values of the material parameters of GaN given
above, we find $V_{0}=0.139$~V and $f_{2}/6f_{1}=-0.988$. Taking
$f_{2}/6f_{1}=-1$, Eq.~(\ref{eq:6}) simplies to 
\begin{equation}
V(x,y,z)\approx-V_{0}\frac{yz^{2}}{r^{3}}.\label{eq:7}
\end{equation}
The piezoelectric field is given by $\boldsymbol{E}=-\mathrm{grad}\,V$.
Since the exciton lifetime depends on the magnitude of the electric
field $E=|\boldsymbol{E}|$ rather than its direction, we visualize
the field distribution in Fig.~\ref{fig:ElectricField} by plotting
an iso-field surface. The shape of this surface does not depend on
$E$ and scales with its value, since $E\propto r^{-1}$. The three-dimensional
surface of constant electric field at $E=10$~kV/cm shown in the
figure extends to distances of about 70~nm from the dislocation line,
and to a depth up to 140~nm from the surface.

The dissociation rate $1/\tau_{E}$ of the free exciton in the electric
field $E$ is calculated analogously to the ionization probability
(per unit time) of the hydrogen atom.\cite{landau:quantmech-IonizationRate,yamabe77,banavar79}
It can be expressed as
\begin{equation}
\frac{1}{\tau_{E}}=\omega\frac{4E_{0}}{E}\exp\left(-\frac{2E_{0}}{3E}\right),\label{eq:8}
\end{equation}
where the frequency $\omega$ and the electric field $E_{0}$ are
given in terms of the exciton binding energy $\mathcal{E}$ as $\omega=\mathcal{E}/\hbar$
and $E_{0}=2\mathcal{E}/e_{0}a_{B}$. Here $\hbar$ is the reduced
Planck constant, $a_{B}=4\pi\kappa_{0}\kappa\hbar^{2}/e_{0}^{2}\mu$
is the Bohr radius, $\mu$ is the reduced mass of the exciton (equal
to 0.18 electron masses for the free \emph{A} exciton in GaN). The
exciton binding energy is $\mathcal{E}=\hbar^{2}/2\mu a_{B}^{2}$.
We thus obtain $e_{0}\mathcal{E}=27$~meV, $\omega=4.1\times10^{13}$~s$^{-1}$,
and $E_{0}=193$~kV/cm. As an example, for the electric field $E=10$~kV/cm
shown in Fig.~\ref{fig:ElectricField}, the characteristic dissociation
time of the exciton is $\tau_{E}=0.12$~ns, i.\,e., notably smaller
than the typical effective exciton lifetime in epitaxial GaN layers.
Since electrons and holes are rapidly separated by the piezoelectric
field, and one carrier type will be driven toward the dislocation,
this process constitutes an effectively nonradiative decay channel
by dissociating excitons already at distances of about 100~nm from
the dislocation line. Equation (\ref{eq:8}) is derived for small
electric fields such that $E\ll E_{0}$. This condition does not restrict
our considerations since, when $E$ becomes comparable with $E_{0}$,
the lifetime $\tau_{E}$ is so small that the dissociation process
can be considered as being instantaneous.

\begin{figure}
\includegraphics[width=1\columnwidth]{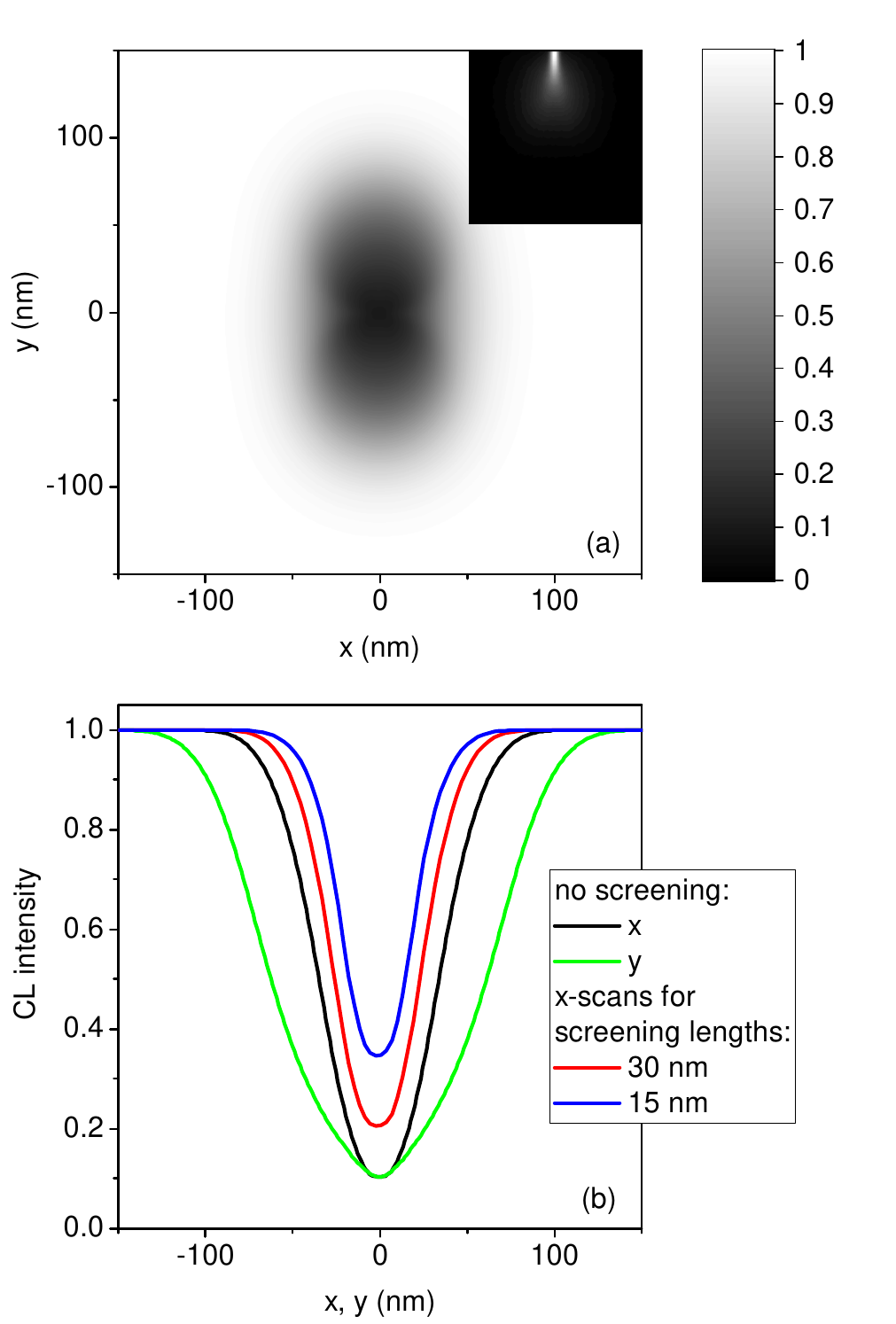}

\caption{Calculated CL intensity map around an edge threading dislocation in
GaN\{0001\} in the absence of exciton diffusion. The acceleration
voltage of the electron beam generating electron-hole pairs is 3~kV.
(a) Two-dimensional CL intensity distribution and (b) scans parallel
($x$ direction) and perpendicular ($y$ direction) to the Burgers
vector of the dislocation. The inset in (a) shows the distribution
of the electron-hole pairs generated by the electron beam on the same
scale as the CL intensity map. The map in (a) is calculated for an
unscreened piezoelectric charge distribution. The profile along $x$
in (b) is compared with the ones calculated for the piezoelectric
charges screened by free carriers with densities $6\times10^{16}$
and $1.5\times10^{16}$~cm$^{-3}$, resulting in room temperature
Debye lengths of 15 and 30~nm, respectively.}

\label{fig:intensity}
\end{figure}

To demonstrate the effect of the electric field displayed in Fig.~\ref{fig:ElectricField}
on exciton recombination in the absence of exciton diffusion, we calculate
a CL image around an edge threading dislocation intersecting the GaN\{0001\}
surface. Electron-hole pairs are generated by the primary electron
beam with an acceleration voltage of 3~kV. The spatial distribution
$Q(\boldsymbol{r})$ of these pairs, which is assumed to correspond
to the distribution of free \emph{A} excitons with wavevector $\boldsymbol{K}=0$,
is obtained with the help of the free software CASINO.\cite{drouin07}
The effective exciton lifetime including exciton dissociation by the
piezoelectric field is calculated according to
\begin{equation}
\frac{1}{\tau_{\mathrm{eff}}(\boldsymbol{r})}=\frac{1}{\tau_{r}}+\frac{1}{\tau_{nr}}+\frac{1}{\tau_{E}(\boldsymbol{r})},\label{eq:9}
\end{equation}
where $\tau_{r}$ and $\tau_{nr}$ are the radiative and nonradiative
lifetimes far from the dislocation, respectively, and $\tau_{E}(\boldsymbol{r})$
is given by Eq.~(\ref{eq:8}). We take $\tau_{r}=200$~ns and $\tau_{nr}=1$~ns,
neglecting any dependence of these values on the electric field. The
CL intensity is calculated then as a convolution
\begin{equation}
I_{\mathrm{CL}}(\boldsymbol{r})=\tau_{r}^{-1}\int\tau_{\mathrm{eff}}(\boldsymbol{r}')Q(\boldsymbol{r}-\boldsymbol{r}')d\boldsymbol{r}'.\label{eq:10}
\end{equation}
The calculated CL map and the intensity profiles along ($x$ direction)
and perpendicular to ($y$ direction) the Burgers vector of the dislocation
are presented in Figs\@.~\ref{fig:intensity}(a) and \ref{fig:intensity}(b).
Longer nonradiative lifetimes would produce profiles with somewhat
larger width. The dependence of the width on the lifetime is, however,
rather weak since $\tau_{E}(\boldsymbol{r})$ sharply varies with
the distance from dislocation.

The inset in Fig.~\ref{fig:intensity}(a) shows, on the same scale
as the CL map, the spatial distribution $Q(\boldsymbol{r})$ of the
electron-hole pairs generated by the electron beam. The figure shows
the projection of the three-dimensional distribution on the $xz$
plane. The majority of electron-hole pairs is generated in a region
with a lateral width of only a few nm, so that the spatial extension
of the dislocation image in Fig.~\ref{fig:intensity} is entirely
determined by the piezoelectric field. This width of about 100~nm
is on the same order as the one observed in experimentally recorded
CL images.\cite{rosner97,sugahara98,speck99,shmidt02,nakaij05,pauc06,yakimov07,ino08,yakimov10,yakimov15,sabelfeld17CL}
The image is notably extended in the $y$ direction (the direction
of the extra half-plane of the dislocation). This asymmetry is expected
to be smoothed out, at least partially, by diffusion of excitons.

Unintentionally doped GaN layers usually exhibit an $n$-type background
doping due to the incorporation of the shallow donors O and Si. These
free electrons will screen the fields induced by the piezoelectric
polarization charges. We follow the Debye-Hückel approximation for
piezoelectric semiconductors\cite{merten66,faivre72,shintani91} and
include in the Poisson equation (\ref{eq:5}) an additional free charge
distribution $\tilde{\varrho}(\boldsymbol{r})=ne_{0}\exp(e_{0}V/k_{B}T)$,
where $n$ is the electron density, $k_{B}$ is the Boltzmann constant
and $T$ is the temperature. Then, after expansion over $e_{0}V\ll k_{B}T$
up to the linear term, Eq.~(\ref{eq:5}) is replaced with 
\begin{equation}
\nabla^{2}V-k_{s}^{2}V=-\frac{\varrho}{\kappa_{0}\kappa},\label{eq:11}
\end{equation}
where the inverse screening length $k_{s}$ is given by $k_{s}=\left(e_{0}^{2}n/\kappa_{0}\kappa k_{B}T\right)^{1/2}$.
For an electron density $n=6\times10^{16}$~cm$^{-3}$ at room temperature,
the Debye screening length is $l_{s}=1/k_{s}=15$~nm. 

The solution of Eq.~(\ref{eq:11}) for a point charge, that ensures
the equipotential surface $V\left|_{z=0}\right.=0$, is the screened
Coulomb potential $\exp(-k_{s}r)/4\pi\kappa_{0}\kappa r$ plus that
of its image (of the opposite sign) with respect to the surface $z=0$.
To calculate the convolution integral of the charge distribution (\ref{eq:2})
with the point charge solution, we represent the charge distribution
(\ref{eq:2}) as a Fourier integral
\begin{equation}
\varrho(x,y,z)=\frac{y}{r_{\perp}}\intop_{0}^{\infty}q\,dq\,J_{1}\left(qr_{\perp}\right)\left(f_{1}+\frac{f_{2}}{3}qz\right)e^{-qz}\label{eq:12}
\end{equation}
and arrive at the potential

\begin{eqnarray}
V & = & \frac{2V_{0}}{k_{s}^{2}}\frac{y}{r_{\perp}}\intop_{0}^{\infty}q\,dq\,J_{1}\left(qr_{\perp}\right)\label{eq:13}\\
 & \times & \left[\left(1-\frac{2f_{2}}{3f_{1}}\frac{q^{2}}{k_{s}^{2}}\right)\left(e^{-qz}-e^{-\sqrt{k_{s}^{2}+q^{2}}z}\right)+\frac{f_{2}}{3f_{1}}qze^{-qz}\right].\nonumber 
\end{eqnarray}
This solution can be directly verified by differentiating the potential
(\ref{eq:13}) and checking that the screened Poisson equation (\ref{eq:11})
is satisfied with the charge distribution (\ref{eq:12}). In the limit
$k_{s}\rightarrow0$, the potential (\ref{eq:13}) reduces to Eq.~(\ref{eq:6}).

Figure \ref{fig:intensity}(b) shows that the screening of the piezoelectric
field reduces both the width of the dislocation image and its contrast.
The region of the electric field, large enough to dissociate excitons,
decreases due to screening both laterally and in depth. The reduction
in lateral direction results in a narrowing of the dislocation image.
Its reduction in depth decreases the probability of nonradiative recombination
for excitons created at larger depths and hence reduces the contrast
of the dislocation image. The effect of the piezoelectric field reduces
with increasing acceleration voltage of the electron beam, since the
piezoelectric field is restricted in depth while the excitons are
produced deeper in the crystal. Then, the nonradiative recombination
at the dislocation core, along the whole dislocation line, becomes
the primary mechanism of the dislocation contrast.

To summarize and conclude, the strain field of an edge threading dislocation
relaxes at the surface to achieve a stress-free boundary. The resulting
strain field causes an inhomogeneous distribution of piezoelectric
polarization charges which, in turn, induces a volume electric field
around the dislocation outcrop. Excitons dissociate in this electric
field, thus reducing the exciton lifetime. Even in the absence of
exciton diffusion, dislocations in undoped GaN give rise to dark spots
in CL maps with diameters up to 100~nm. The Debye-Hückel screening
of the piezoelectric field by free carriers reduces the image diameter.
However, its width remains significant when compared to experimentally
recorded CL intensity profiles of dislocations. Thus, the exciton
diffusion length may be notably smaller than it is usually inferred
from CL images of dislocations. We will study the effect of exciton
diffusion in the presence of the piezofield at the dislocation outcrop
in a forthcoming work.

The authors thank Alexander Tagantsev, Alexander Belov, and Vladimir
Alshits for useful discussions and Uwe Jahn for a critical reading
of the manuscript. K.K.S. acknowledges the support of the Russian
Science Foundation under grant N 14-11-00083.

\bibliographystyle{aipnum4-1}

%

\end{document}